%% file: avs_mlsp.tex
\documentclass{article}
\usepackage{amsmath,graphicx,mlspconf}

\copyrightnotice{978-1-7281-6338-3/21/\$31.00 {\copyright}2021 IEEE}

\toappear{2021 IEEE International Workshop on Machine Learning for Signal Processing, Oct.\ 25--28, 2021, Gold Coast, Australia}

\title{Deep Variational Generative Models\\ for Audio-visual Speech Separation}
\name{Viet-Nhat Nguyen,$^{1,2}$ Mostafa Sadeghi,$^{1,3}$ Elisa Ricci,$^{2}$ and Xavier Alameda-Pineda,$^{1}$}
\address{$^{1}$ RobotLearn Team at Inria Grenoble Rh\^{o}ne-Alpes, France.\\$^{2}$ Universit\`a degli Studi di Trento, Italy.\\$^{3}$Multispeech Team at Inria Nancy Grand-Est}

\usepackage{booktabs}

\usepackage{amsmath,epsfig,amssymb,amsthm,url,amsfonts}
\usepackage{algorithm}% http://ctan.org/pkg/algorithm
\usepackage{algpseudocode}% http://ctan.org/pkg/algorithmicx
\usepackage{multirow}
\usepackage{mdframed}
\usepackage{url}
\usepackage{enumitem}
\usepackage[makeroom]{cancel}
\usepackage{dblfloatfix}
\usepackage{array}
\usepackage{epstopdf}
\usepackage{float}
\usepackage{subfig}
\usepackage{breqn}
\usepackage{cite}
\usepackage{xcolor}
\usepackage{tabularx}
\usepackage[printonlyused,withpage]{acronym}
\usepackage{balance}

\usepackage{makecell}
\usepackage[pagebackref=true,breaklinks=true,letterpaper=true,colorlinks,bookmarks=false]{hyperref}

\setlist[itemize]{label=$\triangleright$}
\newtheoremstyle{break}% name
{}%         Space above, empty = `usual value'
{}%         Space below
{\itshape}% Body font
{}%         Indent amount (empty = no indent, \parindent = para indent)
{\bfseries}% Thm head font
{.}%        Punctuation after thm head
{\newline}% Space after thm head: \newline = linebreak
{}%         Thm head spec
\theoremstyle{break}

\theoremstyle{definition}

\DeclareMathOperator{\vect}{Vect}

\newcommand{\E}{\mathbb{E}}
\newcommand{\bs}{\boldsymbol}

\def\mbb#1{\mathbb{#1}}
\def\bs#1{\boldsymbol{#1}}

\makeatletter
\def\thmhead@plain#1#2#3{%
	\thmname{#1}\thmnumber{\@ifnotempty{#1}{ }\@upn{#2}}%
	\thmnote{ {\the\thm@notefont#3}}}
\let\thmhead\thmhead@plain
\makeatother

\graphicspath{{./figs/}}

\input{Header}

\acrodef{SE}{speech enhancement}
\acrodef{STFT}{short-time Fourier transform}
\acrodef{PSD}{power spectral density}
\acrodef{NMF}{nonnegative matrix factorization}
\acrodef{AV}{audio-visual}
\acrodef{DNN}{deep neural network}
\acrodefplural{DNNs}{deep neural networks}
\acrodef{VAE}{variational auto-encoder}
\acrodefplural{VAEs}{variational auto-encoders}
\acrodef{CVAE}{conditional variational auto-encoder}
\acrodefplural{CVAEs}{conditional variational auto-encoders}
\acrodef{A-VAE}{audio VAE}
\acrodef{V-VAE}{visual VAE}
\acrodef{AV-CVAE}{audio-visual CVAE}
\acrodef{ROI}{region of interest}
\acrodef{MCMC}{Markov Chain Monte Carlo}
\acrodef{EM}{expectation-maximization}
\acrodef{MCEM}{Monte Carlo expectation-maximization}
\acrodef{TF}{time frequency}
\acrodef{ELBO}{evidence lower bound}
\acrodef{ROI}{region of interest}
\acrodef{LR}{Living Room}
\acrodef{SDR}{signal-to-distortion ratio}
\acrodef{PESQ}{perceptual evaluation of speech quality}
\acrodef{ASE}{audio speech enhancement}
\acrodef{VSE}{visual speech enhancement}
\acrodef{AVSE}{audio-visual speech enhancement}
\acrodef{SNR}{signal-to-noise ratio}
\acrodefplural{SNRs}{signal-to-noise ratios}
\acrodef{LSTM}{long short-term memory}
\acrodef{DNNs}{deep neural networks}

\begin{document}

\maketitle

\begin{abstract}
In this paper, we are interested in audio-visual speech separation given a single-channel audio recording as well as visual information (lips movements) associated with each speaker. We propose an unsupervised technique based on audio-visual generative modeling of clean speech. More specifically, during training, a latent variable generative model is learned from clean speech spectra using a variational auto-encoder (VAE). To better utilize the visual information, the posteriors of the latent variables are inferred from mixed speech (instead of clean speech) as well as the visual data. The visual modality also serves as a prior for latent variables, through a visual network. At test time, the learned generative model (both for speaker-independent and speaker-dependent scenarios) is combined with an unsupervised non-negative matrix factorization (NMF) variance model for background noise. All the latent variables and noise parameters are then estimated by a Monte Carlo expectation-maximization algorithm. Our experiments show that the proposed unsupervised VAE-based method yields better separation performance than NMF-based approaches as well as a supervised deep learning-based technique.
\end{abstract}
\vspace{-2mm}

\section{Introduction}
Speech (or speaker) separation, sometimes referred to as the ``cocktail party problem'' \cite{The_Cocktail_Party_Problem}, is the problem of separating speech sources from a mixture involving several speakers. This is a fundamental
task in signal processing with a wide range of applications. In the literature, traditional methods are typically audio-only, for instance based on \ac{NMF} \cite{NMF_SS_Lee2000, NMF_SS_book} or independent component analysis (ICA) \cite{ICA_SS}. Given that the visual modality can provide very useful information, one important question is how to jointly exploit audio and visual data.
%These methods, derived from classical signal processing discipline, primarily rely on matrix factorization. 
%However, the lack of representation power and the inability to easily accommodate other source of information (e.g.\ visual data) limit their applicability to AV speech separation.
% They are not expressive enough to model speech and are difficult to incorporate additional information, e.g. visual data. 

% This is limiting, since the visual modality provides useful information about unknown speech and when efficiently combined with the audio modality, it can help separate different speech sources. With the emergence of deep learning, 
Audio-visual speech separation has been recently addressed in the framework of \ac{DNN}. For example, speech enhancement, which is a special case of speech separation, has been successfully addressed in \cite{gabbay2018visual}, where a neural network model is trained to map noisy audio and the associated visual data to clean audio data. Thus, the visual modality plays an important role allowing to separate speech by treating all the undesired speech sources as noise. Some carefully designed neural networks architectures are also proposed in \cite{Face_Landmark_based_Speaker_independent_Audio_visual_Speech_Enhancement_in_Multi_talker_Environments} to learn a time-frequency mask, exploiting an additional face landmark detector trained on a separate image dataset. However, the second speech considered as the only source of noise is shared among training set and test set. Perhaps the most similar works to the problem we are dealing with are~\cite{Looking_to_Listen_Audio_Visual_Speech_Separation,gao2021visualvoice}. Nevertheless, such methods rely on a huge dataset of synchronized audio-visual data including thousands of hours of video segments from the Web. These DNN-based approaches are versatile to integrate visual information into their architecture. However, they are subject to poor generalization to unseen noise types and require a highly diverse training noise dataset.

Recently, a family of unsupervised methods based on VAEs has been developed to address the speech enhancement problem \cite{ass_vae_Bando_2017, Leglaive_SE2018, AVSE_Using_CVAE,robse_vae_Sadeghi_2020,mixvae_se_Sadeghi_2020}. They rely on deep latent variable generative models consisting of two main steps. First, a generative model is trained on clean speech to learn a probabilistic mapping from latent space to the space of (clean) speech spectrogram. This is done with a VAE network equipped with an encoder and a decoder. The encoder approximates the intractable posterior of the latent variables using a Gaussian distribution. The decoder then acts as a generative model by attempting to reconstruct the clean spectrogram. Secondly, after being jointly trained with the encoder, the decoder is combined with an unsupervised noise variance model, e.g. NMF, to infer both time-varying loudness (gain values) and the model parameters. To make use of available visual modality, \cite{AVSE_Using_CVAE} proposed to apply a conditional VAE network \cite{CVAE_paper}, where the encoder and decoder are conditioned on the visual information. This approach yields state-of-the-art audio-visual speech enhancement in the unsupervised setting.

\begin{figure*}[t!]
	\vspace*{-0.6cm}
	\centering
	\includegraphics[width=0.8\textwidth, height=4.5cm]{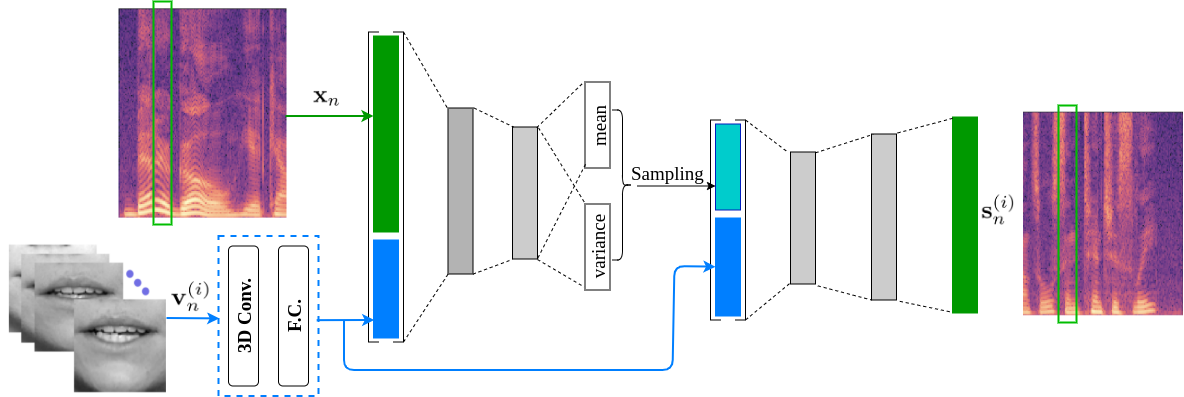}
	\caption{\label{fig:model} VAE model architecture, where $(i)$ indicates speaker number.}
	\vspace*{-0.6cm}
\end{figure*}

In this paper, inspired by previous related literature~\cite{AVSE_Using_CVAE}, we address the single-channel audio-visual speech separation problem. We focus on the situation of two individuals speaking simultaneously in the presence of some background noise, where both faces are available. The main idea is to learn a generative model for clean speech spectrogram (for both speaker-dependent and speaker-independent cases), and combine it with an \ac{NMF} model for the background noise at test time. A Monte-Carlo expectation-maximization framework is developed to estimate the individual speech signals. Our experiments confirm the promising performance of the proposed approach when compared with the traditional \ac{NMF}-based and the recent \ac{DNN}-based techniques. 

The rest of the paper is organized as follows. Section \ref{sec:modelling} presents clean speech modeling based on VAE. Then, Section~\ref{sec:separation} discusses how to separate speech signals  given the learned clean speech model and the observed noisy audio and the associated clean visual data. Finally, the performance of the proposed method is evaluated in Section \ref{sec:experiment}.

%%This section is to present how we integrate visual information to our VAE, how to model clean speeches as a complex gaussian of latent codes
\section{Clean Speech Modelling}
\label{sec:modelling}
\subsection{Model}
We use the same model as in \cite{AVSE_Using_CVAE} to combine the clean speech signal with the visual information. Let $(\bs{s}, \bs{v}) = \{\bs{s}_n \in \mathbb{C}^F, \bs{v}_n \in \mathbb{R}^M\}_{n=1}^N$ denote our observed training set of $N$ synchronized frames of audio and video data where $\bs{s}_n$ represents coefficients of \ac{STFT} at time $n$, and $\bs{v}_n$ is a lip \ac{ROI} embedding. We assume that $\bs{s}$ is generated by a corresponding sequence of latent vectors $\bs{z} = \{ \bs{z}_n \in \mathbb{R}^L \}_{n=1}^N$. Both clean speech $\bs{s}$ and latent code $\bs{z}$ are conditioned on visual features which serve as a deterministic information. Specifically, we have the following latent variable model:
\begin{align}
\label{decoder_VAE}
\bs{s}_n | \mathbf{z}_n, \bs{v}_n;\Theta &\sim \mathcal{N}_c\Big(\boldsymbol{0}, \text{diag}\Big(\bs{\sigma}_{s}(\mathbf{z}_n, \bs{v}_n)\Big)\Big), \\
\label{prior_VAE}
\mathbf{z}_n | \bs{v}_n; \Gamma&\sim \mathcal{N}\Big(\bs{\mu}_p(\bs{v}_n), \text{diag}\Big(\bs{\sigma}_{p}(\bs{v}_n)\Big)\Big), 
\end{align}
where $\mathcal{N}(0, \sigma)$ and $\mathcal{N}_c(0, \sigma)$ denote, respectively, Gaussian and complex proper Gaussian distributions with zero mean and variance $\sigma$. Moreover, $ \bs{\sigma}_{s}(.,.): \mbb{R}^L\times \mbb{R}^M \mapsto 
\mbb{R}_+^F $ is a neural network with parameters denoted $\Theta$. Similarly, $\bs{\mu}_{p}(.) : \mbb{R}^M \mapsto 
\mbb{R}^L$ and $\bs{\sigma}_{p}(.) : \mbb{R}^M \mapsto \mbb{R}_+^L$ are neural networks with parameters $\Gamma$.
\subsection{Training}
To learn the set of parameters $\Theta$ and $\Gamma$, we follow the principles of VAE. In particular, the posterior of the latent codes is approximated by a parameterized Gaussian distribution:
\begin{equation}
\label{eq:varapp}
 q(\bs{z}_n |\bs{s}_n, \bs{v}_n;\Psi)= \mathcal{N}\Big(\bs{\mu}_z(\bs{s}_n, \bs{v}_n), \text{diag}\Big(\bs{\sigma}_{z}(\bs{s}_n, \bs{v}_n)\Big)\Big),
\end{equation}
where, $ \bs{\mu}_z(.,.): \mbb{R}_{+}^F\times \mbb{R}^M \mapsto 
\mbb{R}^L $ and $ \bs{\sigma}_{z}(.,.): \mbb{R}_{+}^F\times \mbb{R}^M \mapsto 
\mbb{R}_+^L $ are neural networks, with parameters denoted $ \Psi $, taking $ \tilde{\mathbf{s}}_n \triangleq (|s_{0n}|^2 \dots |s_{F-1\: n}|^2)^{\top}$ 
as input. To learn all the parameters, the following loss function is optimized \cite{AVSE_Using_CVAE}:
%\label{eq:loss_func}
%\mathcal{L}(\mathbf{x}, &\mathbf{s}^{(1)}, \mathbf{s}^{(2)}, \mathbf{v}^{(1)}, \mathbf{v}^{(1)}; \Theta, \Gamma) = \nonumber \\
%\sum_{i\in\{1,2\}}
%&\Big(
%\alpha \: \mathbb{E}_{q(\mathbf{z}^{(i)}|\mathbf{x}, \mathbf{v}^{(i)}; \Theta)} \left[ \log p(\mathbf{s}^{(i)}|\mathbf{z}^{(i)}, \mathbf{v}^{(i)};\Theta) \right] \nonumber \\
%&+ (1-\alpha) \mathbb{E}_{p(\mathbf{z}^{(i)}|\mathbf{v}^{(i)}; \Gamma)} \left[ \log p(\mathbf{s}^{(i)}|\mathbf{z}^{(i)}, \mathbf{v}^{(i)};\Theta) \right]
%\nonumber \\
%&+ D_{\mathrm{KL}}\left(q(\mathbf{z}^{(i)}|\mathbf{x}, \mathbf{v}^{(i)}; \Theta) \: \| \: p(\mathbf{z}^{(i)}|\mathbf{v}^{(i)}; \Gamma)\right)
%\Big),
%\end{align}
\begin{multline}
\label{eq:loss_func}
\mathcal{L}(\Theta, \Gamma, \Psi) = 
\alpha \: \mathbb{E}_{q(\mathbf{z}|\mathbf{x}, \mathbf{v}; \Psi)} \Big[ \log p(\mathbf{s}|\mathbf{z}, \mathbf{v};\Theta) \Big] + \\
 (1-\alpha) \mathbb{E}_{p(\mathbf{z}|\mathbf{v}; \Gamma)} \Big[ \log p(\mathbf{s}|\mathbf{z}, \mathbf{v};\Theta) \Big] \\
+ \alpha D_{\mathrm{KL}}\Big(q(\mathbf{z}|\mathbf{x}, \mathbf{v}; \Psi) \: \| \: p(\mathbf{z}|\mathbf{v}; \Gamma)\Big),
\end{multline}
in which, $ D_{\mathrm{KL}}(.,.)$ denotes the Kullback-Leibler (KL). Moreover, $\alpha \in [0, 1]$ controls the contributions of the reconstruction errors computed using the latent code sampled from the encoder and the one sampled from the prior network. This strategy takes into account both the reconstruction from audio-visual and visual-only information, which was originally proposed in \cite{AVSE_Using_CVAE} and has been proven effective for speech enhancement. The loss function can be straightforwardly optimized via the reparametrization trick and the gradient descent algorithm as in standard VAEs. In order to better utilize the visual modality during training, noisy speech (instead of the clean one) is used as the input of the encoder. The architecture of our VAE model is depicted in Figure~\ref{fig:model}. The encoder-decoder structure allows us to easily derive a speaker-dependent version, in which the generic decoder is substituted by the ones fine-tuned for each individual speaker. Once trained, the decoder(s) are used for speech separation as described in the following.

\section{Speech Separation}
\label{sec:separation}
\subsection{Unsupervised noise model}
We consider an unsupervised NMF-based Gaussian model for the background noise which is simple and proven effective \cite{ass_vae_Bando_2017}. Specifically, the variance of noise spectrogram could be represented as the product of two non-negative matrices $\mathbf{W} \in \mbb{R}_{+}^{F\times K}$ and $\mathbf{H} \in \mbb{R}_{+}^{K\times N}$. $\mathbf{W}$ consists of $K$ spectral basis vectors and $\mathbf{H}$ is a matrix of temporal activations, with $K$ being called the rank of NMF satisfying $K(F + N) \ll FN$. This leads to the following distribution for the $n$-th noise spectrogram time frame, $\bs{b}_n$:
\begin{equation}
\label{eq:noise_model}
\bs{b}_n\sim \mathcal{N}_c\Big(\boldsymbol{0}, \text{diag}\Big(\Wb\bs{h}_n\Big)\Big),
\end{equation}
where, $ \bs{h}_n $ denotes the $ n $-th column of $ \Hb $.
\subsection{Mixture model}
We model the observed mixture signal recorded using a single-channel microphone as follows ($n=1,\ldots,N$):
\begin{align}
\label{eq:mixture_model}
\bs{x}_n = \sqrt{g^{(1)}_n} \bs{s}_n^{(1)} + \sqrt{g^{(2)}_n} \bs{s}_n^{(2)} + \bs{b}_n.
\end{align}
where, $\bs{s}_n^{(1)}$ and $\bs{s}_n^{(2)}$ denote clean speech signals from two speakers. As suggested in \cite{Leglaive_SE2018}, the gain parameters
$g^{(1)}_n$ and $g^{(2)}_n$  are frequency-invariant and provide robustness against the loudness variations of the speech signals along time. Using (\ref{decoder_VAE}) and (\ref{eq:noise_model}), and an independence assumption, we have:
\begin{multline}
    \bs{x}_{n} | \bs{z}^{(1)}_{n}, \bs{v}^{(1)}_n, \bs{z}^{(2)}_{n}, \bs{v}^{(2)}_n \sim 
\mathcal{N}_c \Big( 
\boldsymbol{0}, g^{(1)}_n \text{diag}\Big(\bs{\sigma}_{s}(\bs{z}_n^{(1)}, \bs{v}_n^{(1)})\Big) + \\ g^{(2)}_n \text{diag}\Big(\bs{\sigma}_{s}(\bs{z}_n^{(2)}, \bs{v}_n^{(2)})\Big) + \text{diag}(\Wb\bs{h}_n) \Big).
\end{multline}
\subsection{Parameters estimation}
Let us denote $\Phi = \{\mathbf{W}, \mathbf{H}, \mathbf{g}^{(1)}, \mathbf{g}^{(2)}\}$ as the collection of parameters we need to estimate from the observed mixture $\mathbf{x}$ and the visual embeddings $\mathbf{v}^{(1)}$, $\mathbf{v}^{(2)}$. Our ultimate goal is to maximize the log-likelihood of $\mathbf{x}$:
\begin{align}
    \max_{\Phi} \int_{\mathbb{Z}} \log p\left(\mathbf{x}| \mathbf{z}^{(1)}, \mathbf{v}^{(1)}, \mathbf{z}^{(2)}, \mathbf{v}^{(2)}; \Phi \right)\textrm{d}\mathbf{z}^{(1)}\textrm{d}\mathbf{z}^{(2)}.
    \label{eq:log-likelihood}
\end{align}
However, directly optimizing the above expression is not feasible due to the intractability of the integration. As in \cite{ass_vae_Bando_2017}, we develop a Monte Carlo Expectation-Maximization (MCEM) method. The algorithm consists of an expectation (E) step and a maximization (M) step.
\subsubsection{E-step}
In this step, the joint probability needs to be averaged over the posterior probability given the current parameters $\Phi^\star$, leading to the following lower bound of~(\ref{eq:log-likelihood}):
\begin{align}
\label{eq:likelihoodEstep}
\mathbb{E}_{p(\mathbf{\mathbf{z}^{(1)}, \mathbf{z}^{(2)}} | \mathbf{x}, \mathbf{v}^{(1)}, \mathbf{v}^{(2)}; \Phi^\star)} \left[ \log p\left(\mathbf{x}, \mathbf{z}^{(1)}, \mathbf{v}^{(1)}, \mathbf{z}^{(2)}, \mathbf{v}^{(2)}; \Phi \right) \right].
\end{align}
Again, the above term cannot be computed analytically. In the MCEM framework, a sequence of $R$ pairs of latent vectors $\{( \mathbf{z}^{(1,r)}_n, \mathbf{z}^{(2,r)}_n ) \}_{r=1}^R~(\forall n)$  are drawn from the posterior $p\left(\mathbf{z}^{(1)}_n, \mathbf{z}^{(2)}_n | \mathbf{x}_n, \mathbf{v}^{(1)}_n, \mathbf{v}^{(2)}_n; \Phi^\star\right)$ to approximate the expected value in (\ref{eq:likelihoodEstep}). Since a direct sampling from this posterior is difficult, we use the Metropolis–Hastings algorithm \cite{MH_algorithm} which is a Markov Chain Monte Carlo (MCMC) method for obtaining a sequence of random samples from a given distribution. This algorithm starts with an initialization and repeatedly makes a random walk to generate new candidates from a transition kernel. Suppose that the current latent values for the $n$-th audio frame are $\mathbf{z}^{(1)}_n$ and $\mathbf{z}^{(2)}_n$. To generate the candidate samples, we choose a symmetric multivariate normal distribution for the transition kernel as it is commonly used:
\begin{equation}
    \mathbf{z}^{\prime(1)}_n | \mathbf{z}^{(1)}_n; \epsilon^2 \sim \mathcal{N}\left( \mathbf{z}^{(1)}_n , \epsilon^2\mathbf{I} \right),
\end{equation}
and similarly for $\mathbf{z}^{(2)}_n$. The new samples are accepted with the following probability:
\begin{align}
\label{eq:accprob1}
    p_{a} = \mathrm{min}\left(1, \frac{p({\bs{z}^{(1)}_n, \bs{z}^{(2)}_n} | \bs{x}_n, \bs{v}^{(1)}_n, \bs{v}^{(2)}_n; \Phi^\star)}
    {p({\bs{z}^{\prime(1)}_n, \bs{z}^{\prime(2)}_n} | \bs{x}_n, \bs{v}^{(1)}_n, \bs{v}^{(2)}_n; \Phi^\star)} \right).
\end{align}
Assuming independence of the latent codes of the two speakers, the acceptance ratio in (\ref{eq:accprob1}) turns out to be:
\begin{align*}
     \frac{p\left(\bs{x}_n | \bs{z}^{(1)}_n, \bs{v}^{(1)}_n, \bs{z}^{(2)}_n, \bs{v}^{(2)}_n; \Phi^\star \right) 
    p\left( \bs{z}^{(1)}_n|\bs{v}^{(1)}_n \right)  p\left( \bs{z}^{(2)}_n|\bs{v}^{(2)}_n \right) }
    {p\left(\bs{x}_n | \bs{z}^{\prime(1)}_n, \bs{v}^{(1)}_n, \bs{z}^{\prime(2)}_n, \bs{v}^{(2)}_n; \Phi^\star \right) 
    p\left( \bs{z}^{\prime(1)}_n|\bs{v}^{(1)}_n \right)  p\left( \bs{z}^{\prime(2)}_n|\bs{v}^{(2)}_n \right) }.
\end{align*}
\begin{figure*}[t]
    \centering
    \includegraphics[width=0.95\textwidth]{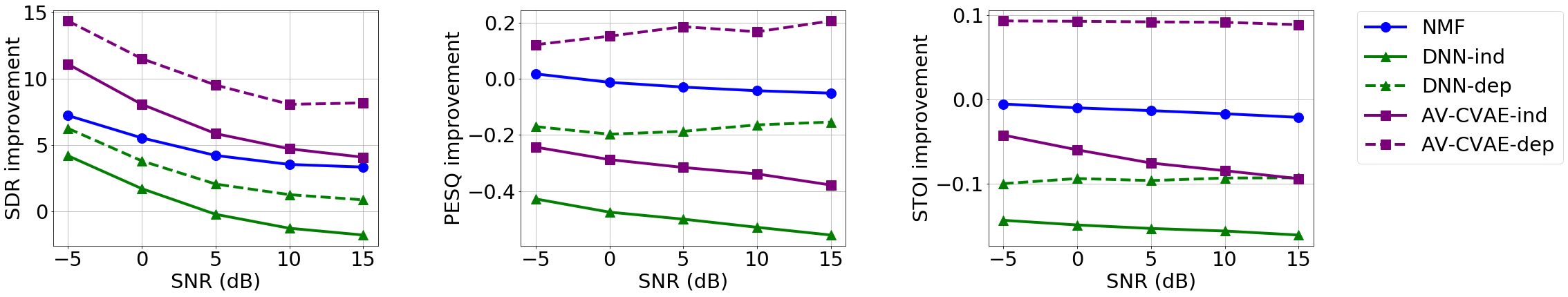}
    \caption{Average measure improvement on the test set: SDR (left), PESQ (center) and STOI (right) as a function of the SNR level. NMF, DNN and CVAE are shown in blue, green and purple respectively. The dashed curves correspond to speaker-dependent models.\vspace{-5mm}}
    \label{fig:results-level}
\end{figure*}
\subsubsection{M-step}
To maximize the log likelihood in (\ref{eq:likelihoodEstep}), we adopt the multiplicative update rule as in \cite{AVSE_Using_CVAE} and generalize it to the case of two clean speech signals:
\begin{align}
    \mathbf{H} & \leftarrow \mathbf{H} \odot 
    \left( 
    \frac{\mathbf{W}^{\mathsf{T}}  \left(|\mathbf{X}|^{\odot 2} \odot \displaystyle \sum_{r=1}^R \left( \mathbf{V}_x^{(r)} \right)^{\odot -2} \right) }
    {\mathbf{W}^{\mathsf{T}} \displaystyle \sum_{r=1}^R \left( \mathbf{V}_x^{(r)} \right)^{\odot -1}}
    \right)^{\odot\frac{1}{2}}, 
    \\
    \mathbf{W} & \leftarrow \mathbf{W} \odot 
    \left( 
    \frac{ \left(|\mathbf{X}|^{\odot 2} \odot \displaystyle \sum_{r=1}^R \left( \mathbf{V}_x^{(r)} \right)^{\odot -2} \right) \mathbf{H}^{\mathsf{T}} }
    {\displaystyle \sum_{r=1}^R \left( \mathbf{V}_x^{(r)} \right)^{\odot -1} \mathbf{H}^{\mathsf{T}} }
    \right)^{\odot\frac{1}{2}},
\end{align}
where $\odot$ denotes entry-wise operations. $\mathbf{V}_x^{(r)}$ is a
matrix whose entries are $g^{(1)}_n \sigma_{s,f}\left(\mathbf{z}^{(1,r)}_n, \mathbf{v}^{(1)}_n \right) + g^{(2)}_n \sigma_{s,f}\left(\mathbf{z}^{(2,r)}_n, \mathbf{v}^{(2)}_n\right) + (\mathbf{W} \mathbf{H})_{fn}$,~$\forall f,n$.  In a similar fashion, the formula to update the vector of speech gains is as follows:
\begin{multline}
\label{eq:update_gains}
    \left(\mathbf{g}^{(i)}\right)^{\mathsf{T}} \leftarrow \left(\mathbf{g}^{(i)}\right)^{\mathsf{T}} \odot
    \\
    \left( 
    \frac{\mathbf{1}^{\mathsf{T}} 
        \left( 
        |\mathbf{X}|^{\odot 2} \odot \displaystyle \sum_{r=1}^R \left( 
                        \mathbf{V}_s^{(i,r)} \odot \left( \mathbf{V}_x^{(r)} \right)^{\odot-2}
                     \right) 
        \right)
        }
        {
        \mathbf{1}^{\mathsf{T}} \left( 
        \sum_{r=1}^R \left( 
                        \mathbf{V}_s^{(i,r)} \odot \left( \mathbf{V}_x^{(r)} \right)^{\odot-1}
                     \right) 
        \right)
        }
    \right)^{\odot\frac{1}{2}},
\end{multline}
in which $i \in \{1, 2\}$ indicates speaker number and $\mathbf{1}$ denotes a row vector of size $F$ with all the elements being $1$. Similarly, $\mathbf{V}_s^{(i,r)}$ consists of entries being $\sigma_{s,f}\left(\mathbf{z}^{(i,r)}_n, \mathbf{v}^{(i)}_n \right)$.\vspace{-2mm}
\subsection{Speech separation}
Having estimated the parameters $\hat{\Phi} = \{ \hat{\mathbf{W}}, \Hat{\mathbf{H}}, \hat{\mathbf{g}}^{(1)}, \hat{\mathbf{g}}^{(2)} \}$, let us denote $\tilde{s}^{(i)}_{fn} = \sqrt{\hat{g}^{(i)}_n}s_{fn}$ as the scaled version of clean speech spectra. Moreover, let
\begin{equation}
	q_n(\bs{z}^{(1)}_n, \bs{z}^{(2)}_n) = p\left( \bs{z}^{(1)}_n, \bs{z}^{(2)}_n \vert \bs{x}_n, \bs{v}^{(1)}_n, \bs{v}^{(2)}_n; \hat{\Phi} \right)
\end{equation}
 denote the posterior of the latent variables. The clean speech STFT coefficients of speaker 1 are estimated in a probabilistic manner as follows:
\begin{align}
\label{eq:est_s1}
\hat{\tilde{s}}^{(1)}_{fn} 
&= \mathbb{E}_{p(\tilde{s}_{fn}^{(1)} \vert \bs{x}_n, \bs{v}^{(1)}_n, \bs{v}^{(2)}_n; \hat{\Phi})}[ \tilde{s}^{(1)}_{fn}] \nonumber \\
&= \mathbb{E}_{q_n(\bs{z}^{(1)}_n, \bs{z}^{(2)}_n)}\left[
   \mathbb{E}_{p(\tilde{s}^{(1)}_{fn} \vert \bs{x}_n, \bs{z}^{(1)}_n, \bs{v}^{(1)}_n, \bs{z}^{(2)}_n, \bs{v}^{(2)}_n; \hat{\Phi})} [ \tilde{s}^{(1)}_{fn} ]
\right].
\end{align}
The posterior of $\tilde{s}^{(1)}_{fn}$ can be simplified as follows:
\begin{align}
\label{eq:posterior_s1}
    &p(\tilde{s}^{(1)}_{fn} \vert \bs{x}_n, \bs{z}^{(1)}_n, \bs{v}^{(1)}_n, \bs{z}^{(2)}_n, \bs{v}^{(2)}_n; \hat{\Phi}) \nonumber \\
    &\propto p(\bs{x}_{fn} \vert \tilde{s}^{(1)}_{fn}, \bs{z}^{(2)}_n, \bs{v}^{(2)}_n; \hat{\Phi}) \times p(\tilde{s}^{(1)}_{fn} | \bs{z}^{(1)}_n, \bs{v}^{(1)}_n) \nonumber \\
    &\sim \mathcal{N}_c\left(x_{fn}; \tilde{s}^{(1)}_{fn}, \hat{g}^{(2)}_n \sigma_{s,f}(\bs{z}^{(2)}_n, \bs{v}^{(2)}_n) + (\hat{\Wb}\hat{\mathbf{H}})_{fn} \right) \times \nonumber \\
    &\;\;\;\;\:\mathcal{N}_c\left(\tilde{s}^{(1)}_{fn}; 0,  \hat{g}^{(1)}_n \sigma_{s,f}(\bs{z}^{(1)}_n, \bs{v}^{(1)}_n) \right).
\end{align}
By combining (\ref{eq:posterior_s1}), (\ref{eq:est_s1}), we obtain the final estimation of clean speech STFT coefficients:
\begin{align}
    \hat{\tilde{s}}^{(1)}_{fn} &= \mathbb{E}_{q_n(\bs{z}^{(1)}_n, \bs{z}^{(2)}_n)}
    \left[
    \frac{\hat{g}^{(1)}_n \sigma_f(\bs{z}^{(1)}_n, \bs{v}^{(1)}_n)}{\hat{V}_f(\bs{z}^{(1)}_n, \bs{z}^{(2)}_n)}
    \right] ,
%    \\
%    \hat{\tilde{s}}^{(2)}_{fn} &= \mathbb{E}_{\mathbf{z}^{(1)}_n, \mathbf{z}^{(2)}_n \sim q_n}
%    \left[
%    \frac{\hat{g}^{(2)}_n \sigma_f(\mathbf{z}^{(2)}_n, \mathbf{v}^{(2)}_n)}{\hat{V}_f(\mathbf{z}^{(1)}_n, \mathbf{z}^{(2)}_n)}
%    \right],
\end{align}
and analogously for $\tilde{s}^{(2)}_{fn}$. Here, $\hat{V}_f(\mathbf{z}^{(1)}_n, \mathbf{z}^{(2)}_n)$ is the mixture variance at the frequency $f$ modeled in (\ref{eq:mixture_model}) with learned parameters $\hat{\Phi}$. These expectations could be approximated using the same MCMC method as before.

\begin{table*}[t]
\centering
    \caption{Average improvement for different configurations.}
    \label{tab:improvement_gender}
    \begin{tabular}{c|ccc|ccc|ccc} % <-- 
    \toprule
    Gender & \multicolumn{3}{c|}{Same (Male)} 
    & \multicolumn{3}{c|}{Different } 
    & \multicolumn{3}{c}{Same (Female)} \\ 
    Measure & SDR & PESQ & STOI & SDR & PESQ & STOI & SDR & PESQ & STOI \\ \midrule
    NMF         & 3.47 & -0.21 & -0.04 & 6.27 & 0.12 & 0.02 & 2.40 & -0.17 & -0.07 \\
    % \hline
    DNN-ind     & -0.95 & -0.49 & -0.16 & 1.62 & -0.45 & -0.15 & -0.47 & -0.64 & -0.16\\
    % \hline
    DNN-dep     & 0.78 & -0.06 & -0.04 & 1.92 & -0.07 & -0.04 & 0.77 & -0.17 & -0.06\\
    % \hline
    AV-CVAE-ind & 5.27 & -0.55 & -0.10 & 7.89 & -0.19 & -0.05 & 5.72 & -0.32 & -0.09\\
    % \hline
    AV-CVAE-dep & 9.43 & 0.01 & 0.10 & 11.27 & 0.29 & 0.10 & 9.47 & 0.09 & 0.06\\
    \bottomrule
    \end{tabular}
\end{table*}
\begin{figure*}[t]
    \centering
    \includegraphics[width=0.95\textwidth]{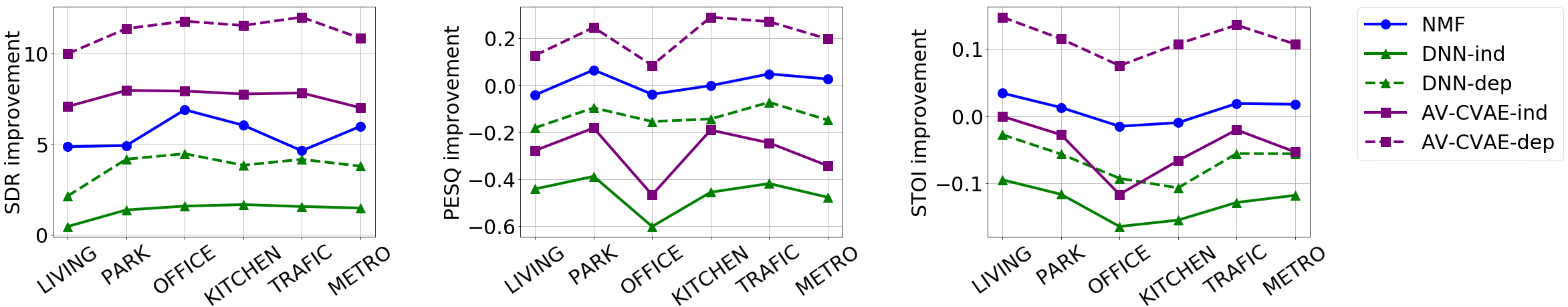}
    \caption{Average measure improvement on the test set: SDR (left), PESQ (center) and STOI (right) per noise type. NMF, DNN and CVAE are shown in blue, green and purple respectively. The dashed curves correspond to speaker-dependent models.\vspace{-5mm}}
    \label{fig:results-type}
\end{figure*}

\section{Experiments}\label{sec:experiment}
\textbf{Dataset.} We evaluate the performance of our method on a mixed-speech version created from the TCD-TIMIT dataset \cite{TCDTIMIT_dataset} which is widely used in \ac{AV} speech processing research. This corpus  contains \ac{AV} recordings of $ 59 $ English speakers uttering $98$ sentences each. Following~\cite{NTCD_TIMIT_dataset}, the visual data are extracted from raw videos by using a joint face and ROI detector. This results in 30~FPS videos of lips \acp{ROI} of size 67$\times$67 pixels. The speech signals are sampled at $16$~kHz and represented by audio spectral features computed using Short-time Fourier transform with a STFT window of 64~ms (hence $F=513$) and 50\% overlap.

The dataset is divided into disjoint sets of $42$, $8$ and $9$ speakers for training, validation and testing, respectively. In each set, mixed speech signals are formed by randomly adding two utterances from two different speakers. Six types of noise taken from the DEMAND dataset \cite{thiemann_joachim_2013_1227121} are added to the mixed speech signals in the test set, with noise levels: $ \lk -15, -10, -5, 0, 5 \rk $~dB, including: DLIVING (living room), NPARK (outside park), OOFFICE (small office), DKITCHEN (inside a kitchen), STRAFFIC (busy traffic intersection) and TMETRO (subway).
% \begin{itemize}
%     \item DLIVING: in a living room, mostly are music playing.
%     \item NPARK: in an outside park, has people walking around, bird sound and leaves rustling.
%     \item OOFFICE: a small office with people using computers.
%     \item DKITCHEN: inside a kitchen during food preparation.
%     \item STRAFFIC: in a busy traffic intersection.
%     \item TMETRO: in a subway, includes passengers announcement voice.
% \end{itemize}
In the speaker-dependent approach, 80 speech signals from each speaker are dedicated to fine-tune the model and the rest is used for evaluating the performance.

\textbf{Baseline methods.} To assess the performance of our proposed method, we compare it with two other methods: an \ac{NMF}-based \cite{NMF_SS_book} approach which is a dictionary learning and speaker-dependent method, and a \ac{DNN}-based \cite{gabbay2018visual} approach trained to directly reconstruct spectra of constituent utterances. While the former is audio-only and semi-supervised, the latter is audio-visual and fully supervised. %The proposed model lies in between.

% The former is an audio-only speaker-dependent approach, which is commonly referred to as a semi-supervised method in the source separation literature. Firstly, a global dictionary is extracted from all of the clean speech signals (including all the speakers). Then, this matrix serves as an initialization to learn a dictionary for each speaker. At test time, only the sub-dictionary corresponding to noise signal is learned whereas the part of the dictionary matrix corresponding to clean speech signals is fixed. Each clean speech is then estimated by Wiener filtering. The method of \cite{gabbay2018visual} makes extensive use of the power of \ac{DNN} to learn \ac{AV}
% features from which a spectrogram of the clean speeche is decoded using an audio decoder. We implemented a speaker-independent as well as a speaker-dependent version of this method.

\textbf{Architecture and parameter settings.}
The architecture of our AV-CVAE is similar to that of \cite{AVSE_Using_CVAE}, which consists of a visual feature extractor and a standard encoder-decoder. In contrast to \cite{AVSE_Using_CVAE} which uses a fully-connected network for the visual feature extractor, we propose to use 3 layers of 3D-convolution, followed by 2 fully-connected layers. The encoder has 2 fully-connected layers outputting a 128-dimensional latent space. The decoder is symmetric. %The code will be made available upon acceptance of the manuscript.

% here, we use convolutional layers. More precisely, the network is composed of three 3D-convolutional layers which takes as the input 3 consecutive raw frames with kernel size of 3 and ReLU activation functions, followed by 2 hidden layers to embed the lip ROIs into a feature vector of length 256. A stride of 2 is used for all the convolutions. The clean spectrogram time frames are combined with the corresponding extracted visual features before passing through an encoder which is responsible for encoding both visual and audio information into a 128-dimensional latent space. Both encoder and decoder have 2 fully connected layers with 256 nodes and hyperbolic tangent activation functions. 

As suggested in \cite{Leglaive_SE2018}, all the dictionaries for clean speech signals have rank $K_{speech} = 64$, yielding the best speech separation performance. Moreover, the rank of the noise dictionary is arbitrarily fixed to $K_{noise} = 10$. To assess the speech separation performance of the method proposed in \cite{gabbay2018visual}, we adapt its available implementation for speech enhancement to speech separation. To this end, a second speech signal is added to the main speech instead of noise, to form a mixed audio. The rest is kept the same.
%entire architecture as well as data pre-processing are kept the same as in \cite{gabbay2018visual}. 
% In particular, the input includes mel-spectrogram calculated from mixture and its associated 5 consecutive visual \ac{ROI}s whose size is $128 \times 128$. The network makes use of convolutional neural network where the video encoder and audio encoder have 6 and 5 convolutional layers with various sizes of kernel and stride. The audio decoder consists of 5 transposed convolution layers which are symmetric with respect to the layers of the audio encoder.

\textbf{Training protocol.}
The proposed VAE architecture is trained with Adam optimizer \cite{KingmaB14} with a constant learning rate of $10^{-4}$ for over 45 epochs as determined by the separation performance in the validation set. %In every epoch, each speech signal is combined with another one from another speaker to form an input audio, in which the shorter signal is padded with zero. 
The trade-off weight $\alpha$ in loss function (\ref{eq:loss_func}) is set to $0.9$. In the speaker-dependent scenario, the VAE is fine-tuned only for 1 epoch to avoid overfitting. The batch size is set to 256.% is set for both training and tuning. %It is worth mentioning the significant impact of the i.i.d. samples in one batch to the final performance. Therefore it is important to have pairs of audio-visual frames taken from different utterances appear in one batch, for the purpose of reducing data correlation. To this end, whilst taking into account the memory limitation of resources, we process and shuffle a portion of training set in one sitting.

\textbf{Results.} We evaluate the speech separation performance in terms of signal-to-distortion ratio (SDR) \cite{SDR_measurement}, the short-time objective intelligibility (STOI) \cite{STOI_measurement}, and the perceptual evaluation of speech quality (PESQ) \cite{PESQ_measurement} as standard metrics. We report all the measurements in terms of improvement compared with the obtained scores evaluated on the unprocessed mixture signal. We report our results as a function of the noise power and the noise type in Figure~\ref{fig:results-level} and in Figure~\ref{fig:results-type} respectively. In terms of SDR, our CVAE-based methods exhibit superior performance compared to the DNN-based counterparts and to the NMF. In terms of PESQ, the proposed unsupervised approach outperforms the DNN-based counterparts, but only the CVAE-based speaker-dependent outperforms the NMF, which is speaker-dependent too. A similar behavior is observed for STOI, with a small difference. In some cases, the speaker independent CVAE outperforms the speaker-dependent DNN, which is quite a remarkable achievement. Similar conclusions are obtained when the methods are compared in different gender configurations, see Table~\ref{tab:improvement_gender}. One clear phenomenon observed is that the two speakers of different gender are easily separated (by all methods) than two of the same gender.
\color{black}

\balance

\section{Conclusions}

In this paper, we presented an unsupervised variational generative modeling framework, inspired by~\cite{AVSE_Using_CVAE}, to address the single-channel \ac{AV} speech separation problem. 
% This is a generalization of the previous work \cite{AVSE_Using_CVAE} originally developed for speech enhancement. 
The proposed method consists of a training phase to learn an audio-visual generative model based on VAE for clean speech, and a test phase to estimate clean speech signals from observed signle-channel audio as well as the visual information. To model the background noise, a variance model based on NMF is used. An MCEM method is developed for estimating noise parameters and clean speech signals. Our experiments showed that the proposed CVAE method exhibits better performance than a DNN-based baseline on the TCD-TIMIT dataset, both for speaker-independent and speaker-dependent scenarios.
% As a future work, one may consider integrating a recurrent VAE as suggested in \cite{RNN-VAE}, or encouraging the correlation of latent variables, to further improve the estimated speech signals' intelligibilities.

% In this paper, we presented an unsupervised variational generative modeling framework to address the single-channel \ac{AV} speech separation problem. This is a generalization of the previous work \cite{AVSE_Using_CVAE} originally developed for speech enhancement. The proposed method consists of a training phase to learn an audio-visual generative model based on VAE for clean speech, and a test phase to estimate clean speech signals from observed signle-channel audio as well as the visual information. To model the background noise, a variance model based on NMF is used. An MCEM method is also developed for estimating noise parameters and clean speech signals.

% Our experiments showed that in unsupervised settings, the proposed AV-CVAE method gave better performance than a DNN-based baseline on the TCD-TIMIT dataset, both for speaker-independent and speaker-dependent scenarios. As a future work, one may consider integrating a recurrent VAE as suggested in \cite{RNN-VAE}, or encouraging the correlation of latent variables, to further improve the estimated speech signals' intelligibilities.

\section{Acknowledgments}
This research was partially funded by the ANR ML3RI (ANR-19-CE33-0008-01) the Multidisciplinary Institute of Artificial Intelligence (ANR-19-P3IA-0003).

\bibliographystyle{IEEEbib}
\bibliography{myref_compressed}

\end{document}

%% file: Header.tex
\newcommand{\lk}{ \left\{ }
\newcommand{\rk}{ \right\} }

\newcommand{\tr}{ {\mbox{{trace}}} }

\newcommand{\Rbb}{{\mathbb{R}}}

\newcommand{\Hb}{{\bf H}}

\newcommand{\bb}{{\bf b}}

\newcommand{\sbb}{{\bf s}}

\newcommand{\xb}{{\bf x}}

\newcommand{\zb}{{\bf z}}

\newcommand{\Wb}{{\bf W}}

\newcommand{\cc}{{\cal c}}

\newcommand{\alphab}{{\mbox{\boldmath $\alpha$}}}

\newcommand{\vb}{{\mathbf v}}

\newsavebox\mybox

%% file: avs_mlsp.bbl
\begin{thebibliography}{10}

\bibitem{The_Cocktail_Party_Problem}
Simon Haykin and Zhe Chen,
\newblock ``The cocktail party problem,''
\newblock {\em Neural Computation}, vol. 17, no. 9, pp. 1875--1902, 2005.

\bibitem{NMF_SS_Lee2000}
Daniel~D. Lee and H.~Sebastian Seung,
\newblock ``Algorithms for non-negative matrix factorization,''
\newblock in {\em NeurIPS}, Cambridge, MA, USA, 2000, pp. 535--541, MIT Press.

\bibitem{NMF_SS_book}
C\'edric F\'evotte, Emmanuel Vincent, and Alexey Ozerov,
\newblock ``Single-channel audio source separation with nmf: divergences,
  constraints and algorithms,''
\newblock in {\em Audio Source Separation}, pp. 1--24. Springer, 2018.

\bibitem{ICA_SS}
{Jen-Tzung Chien} and {Bo-Cheng Chen},
\newblock ``A new independent component analysis for speech recognition and
  separation,''
\newblock {\em IEEE/ACM TASLP}, vol. 14, no. 4, pp. 1245--1254, 2006.

\bibitem{gabbay2018visual}
Aviv Gabbay, Asaph Shamir, and Shmuel Peleg,
\newblock ``Visual speech enhancement,''
\newblock in {\em Interspeech}. 2018, pp. 1170--1174, {ISCA}.

\bibitem{Face_Landmark_based_Speaker_independent_Audio_visual_Speech_Enhancement_in_Multi_talker_Environments}
G.~{Morrone}, S.~{Bergamaschi}, L.~{Pasa}, L.~{Fadiga}, V.~{Tikhanoff}, and
  L.~{Badino},
\newblock ``Face landmark-based speaker-independent audio-visual speech
  enhancement in multi-talker environments,''
\newblock in {\em IEEE ICASSP}, 2019, pp. 6900--6904.

\bibitem{Looking_to_Listen_Audio_Visual_Speech_Separation}
Ariel Ephrat, Inbar Mosseri, Oran Lang, Tali Dekel, Kevin Wilson, Avinatan
  Hassidim, William~T. Freeman, and Michael Rubinstein,
\newblock ``Looking to listen at the cocktail party: A speaker-independent
  audio-visual model for speech separation,''
\newblock {\em ACM Trans. Graph.}, vol. 37, no. 4, July 2018.

\bibitem{gao2021visualvoice}
Ruohan Gao and Kristen Grauman,
\newblock ``Visualvoice: Audio-visual speech separation with cross-modal
  consistency,''
\newblock in {\em IEEE/CVF CVPR}, 2021.

\bibitem{ass_vae_Bando_2017}
Y.~{Bando}, M.~{Mimura}, K.~{Itoyama}, K.~{Yoshii}, and T.~{Kawahara},
\newblock ``Statistical speech enhancement based on probabilistic integration
  of variational autoencoder and non-negative matrix factorization,''
\newblock in {\em IEEE ICASSP}, 2018, pp. 716--720.

\bibitem{Leglaive_SE2018}
S.~{Leglaive}, L.~{Girin}, and R.~{Horaud},
\newblock ``A variance modeling framework based on variational autoencoders for
  speech enhancement,''
\newblock in {\em IEEE MLSP}, 2018, pp. 1--6.

\bibitem{AVSE_Using_CVAE}
M.~{Sadeghi}, S.~{Leglaive}, X.~{Alameda-Pineda}, L.~{Girin}, and R.~{Horaud},
\newblock ``Audio-visual speech enhancement using conditional variational
  auto-encoders,''
\newblock {\em IEEE/ACM TASLP}, vol. 28, pp. 1788--1800, 2020.

\bibitem{robse_vae_Sadeghi_2020}
M.~{Sadeghi} and X.~{Alameda-Pineda},
\newblock ``Robust unsupervised audio-visual speech enhancement using a mixture
  of variational autoencoders,''
\newblock in {\em IEEE International Conference on Acoustics, Speech and Signal
  Processing (ICASSP)}, 2020.

\bibitem{mixvae_se_Sadeghi_2020}
M.~{Sadeghi} and X.~{Alameda-Pineda},
\newblock ``Mixture of inference networks for vae-based audio-visual speech
  enhancement,''
\newblock {\em IEEE TSP}, 2021.

\bibitem{CVAE_paper}
Kihyuk Sohn, Honglak Lee, and Xinchen Yan,
\newblock ``Learning structured output representation using deep conditional
  generative models,''
\newblock in {\em NeurIPS}, C.~Cortes, N.~D. Lawrence, D.~D. Lee, M.~Sugiyama,
  and R.~Garnett, Eds., pp. 3483--3491. Curran Associates, Inc., 2015.

\bibitem{MH_algorithm}
W.~K. Hastings,
\newblock ``{Monte Carlo sampling methods using Markov chains and their
  applications},''
\newblock {\em Biometrika}, vol. 57, no. 1, pp. 97--109, 04 1970.

\bibitem{TCDTIMIT_dataset}
N.~{Harte} and E.~{Gillen},
\newblock ``Tcd-timit: An audio-visual corpus of continuous speech,''
\newblock {\em IEEE TMM}, vol. 17, no. 5, pp. 603--615, 2015.

\bibitem{NTCD_TIMIT_dataset}
Ahmed Hussen~Abdelaziz,
\newblock ``Ntcd-timit,'' May 2017.

\bibitem{thiemann_joachim_2013_1227121}
J.~Thiemann, N.~Ito, and E.~Vincent,
\newblock ``{DEMAND: a collection of multi-channel recordings of acoustic noise
  in diverse environments},'' 2013.

\bibitem{KingmaB14}
Diederik~P. Kingma and Jimmy Ba,
\newblock ``Adam: {A} method for stochastic optimization,''
\newblock in {\em ICLR}, 2015.

\bibitem{SDR_measurement}
E.~{Vincent}, R.~{Gribonval}, and C.~{Fevotte},
\newblock ``Performance measurement in blind audio source separation,''
\newblock {\em IEEE/ACM TASLP}, vol. 14, no. 4, pp. 1462--1469, 2006.

\bibitem{STOI_measurement}
C.~H. {Taal}, R.~C. {Hendriks}, R.~{Heusdens}, and J.~{Jensen},
\newblock ``A short-time objective intelligibility measure for time-frequency
  weighted noisy speech,''
\newblock in {\em IEEE ICASSP}, 2010, pp. 4214--4217.

\bibitem{PESQ_measurement}
A.~W. {Rix}, J.~G. {Beerends}, M.~P. {Hollier}, and A.~P. {Hekstra},
\newblock ``Perceptual evaluation of speech quality (pesq)-a new method for
  speech quality assessment of telephone networks and codecs,''
\newblock in {\em IEEE ICASSP}, 2001.

\end{thebibliography}
